\pdfoutput=1

\documentclass[conference]{IEEEtran}
\usepackage{ifpdf}

\ifCLASSINFOpdf
 \usepackage[pdftex]{graphicx}
 \graphicspath{{figures/}}
 \DeclareGraphicsExtensions{.pdf,.jpeg,.png}
\else
 \usepackage[dvips]{graphicx}
 \graphicspath{{../eps/}}
 \DeclareGraphicsExtensions{.eps}
\fi
\usepackage{fixltx2e}

\usepackage{stfloats}

\usepackage{url}

\usepackage{hyperref}
\hypersetup{
  pdfauthor={Bernhard Haslhofer, Rainer Simon, Robert Sanderson, Herbert Van De Sompel},
  pdftitle={The Open Annotation Collaboration (OAC) Model}}

\usepackage{color}

\begin{document}
%
\title{The Open Annotation Collaboration (OAC) Model}



%
\author{\IEEEauthorblockN{Bernhard Haslhofer\IEEEauthorrefmark{1},
Rainer Simon\IEEEauthorrefmark{2},
Robert Sanderson\IEEEauthorrefmark{3} and
Herbert van de Sompel\IEEEauthorrefmark{3}}
\IEEEauthorblockA{\IEEEauthorrefmark{1}Cornell Information Science, Ithaca, USA\\
\emph{bernhard.haslhofer@cornell.edu}}
\IEEEauthorblockA{\IEEEauthorrefmark{2}Austrian Institute of Technology, Vienna, Austria\\
\emph{rainer.simon@ait.ac.at}}
\IEEEauthorblockA{\IEEEauthorrefmark{3}Los Alamos National Laboratory, Los Alamos, USA\\
\emph{\{rsanderson,herbertv\}@lanl.gov}}}


\maketitle

\begin{abstract}

Annotations allow users to associate additional information with existing resources. Using proprietary and closed systems on the Web, users are already able to annotate multimedia resources such as images, audio and video. So far, however, this information is almost always kept locked up and inaccessible to the Web of Data. We believe that an important step to take is the integration of multimedia annotations and the Linked Data principles. This should allow clients to easily publish and consume, thus exchange annotations about resources via common Web standards.
We first present the current status of the Open Annotation Collaboration, an international initiative that is currently working on annotation interoperability specifications based on best practices from the Linked Data effort. Then we present two use cases and early prototypes that make use of the proposed annotation model and present lessons learned and discuss yet open technical issues.

\end{abstract}



%
\IEEEpeerreviewmaketitle

\section{Introduction}\label{sec:intro}

Large scale media portals such as Youtube and Flickr allow users to attach information to multimedia objects by means of annotations. Web portals hosting multi-lingual collections of millions of digitized items such as Europeana are currently investigating how to integrate the knowledge of end users with existing digital curation processes. Annotations are also becoming an increasingly important component in the cyber-infrastructure of many scholarly disciplines.

A Web-based annotation model should fulfill several requirements. In the age of video blogging
and real-time sharing of geo-located images, the notion of solely textual annotations has become obsolete. Therefore, multimedia Web resources should be annotatable and also be able to be annotated onto other resources. Users often discuss multiple
segments of a resource, or multiple resources, in a single annotation
and thus the model should support multiple targets. An annotation framework should also follow the Linked Open Data guidelines to promote annotation sharing between systems. In order to avoid inaccurate or incorrect annotations, it must take the ephemeral nature of Web resources into account.

Annotations on the Web have many facets: a simple example could be a textual note or a tag (cf.,~\cite{Hunter:2009bh}) annotating an image or video. Things become more complex when a particular paragraph in an HTML document annotates a segment (cf.,~\cite{Hausenblas:2009fk}) in an online video or when someone draws polygon shapes on tiled high-resolution image sets. If we further extend the annotation concept, we could easily regard a large portion of Twitter tweets as annotations on Web resources. Therefore, in a generic and Web-centric conception, we regard an annotation as association created between one \emph{body} resource and other \emph{target} resources, where the body must be somehow \emph{about} the target.

Annotea~\cite{Kahan:2001vn} already defines a specification for publishing annotations but has several shortcomings: (i) it was designed for the annotation of Web pages and provides only limited means to address segments in multimedia objects, (ii) if clients want to access annotations they need to be aware of the Annotea-specific protocol, and (iii) Annotea annotations do not take into account that Web resources are very likely to have different states over time.

Throughout the years several Annotea extensions have been developed to deal with these and other shortcomings: Koivunnen~\cite{Koivunen:2006s} introduced additional types of annotations, such as Bookmark and Topic. Schroeter and Hunter~\cite{Schroeter:uq} proposed to express segments in media-objects by using \emph{context} resources in combination with formalized or standardized descriptions to represent the context, such as SVG or complex datatypes taken from the MPEG-7 standard. Based on that work, Haslhofer et al.~\cite{Haslhofer:2009ve} introduce the notion of \emph{annotation profiles} as containers for content- and annotation-type specific Annotea extensions and suggested that annotations should be dereferencable resources on the Web, which follow the Linked Data principles. However, these extensions were developed separate from each other and inherit the above-mentioned Annotea shortcomings.

In this paper, we describe how the Open Annotations Collaboration (OAC), an effort aimed at establishing annotation interoperability, tackles these issues. We describe two annotation use cases --- image and historic map annotation -- for which we have implemented the OAC model and report on lessons learned and open issues. We also briefly summarize related work in this area and give outlook on our future work.

\section{The Open Annotation Collaboration}\label{sec:oac}

The Open Annotation Collaboration (OAC) is an international group with the aim of providing a Web-centric, interoperable annotation environment that facilitates cross-boundary annotations, allowing multiple servers, clients and overlay services to create, discover and make use of the valuable information contained in annotations. To this end, a Linked Data based data model has been adopted.

\subsection{Open Annotation Data Model}

\begin{figure}[!t]
  \centering
  \includegraphics[width=\columnwidth]{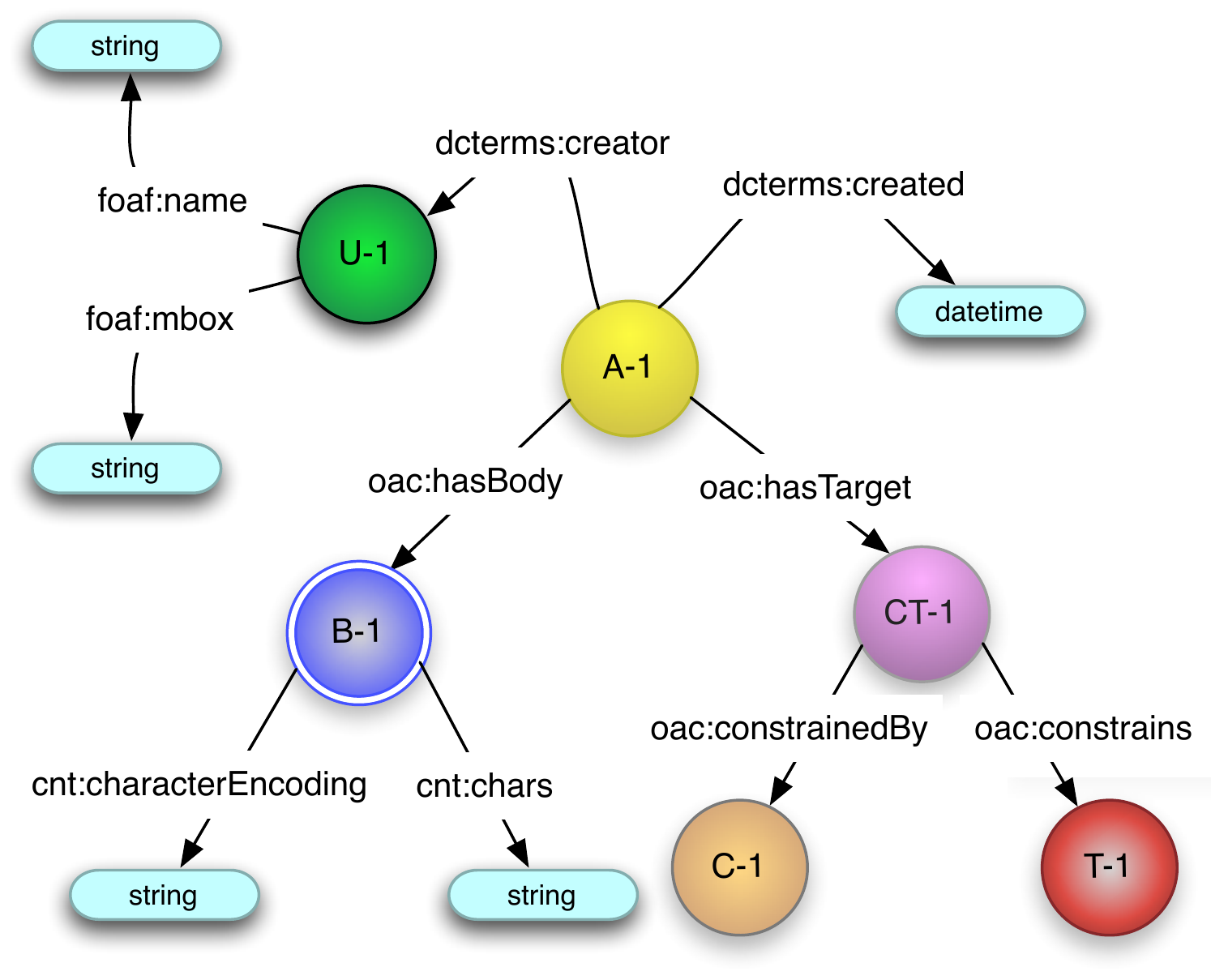}
  \caption{OAC Baseline Data Model.}
\label{fig:oac_data_model}
\end{figure}

The OAC data model tries to pull together various extensions of Annotea into a cohesive whole. The Web architecture and Linked Data guidelines are foundational principles, resulting in a specification that can be applied to annotate any set of Web resources. At the time of this writing, the specification, which is available at \url{http://www.openannotation.org/spec/alpha3/}, is still under development. Following its predecessors, the OAC model, shown in Figure \ref{fig:oac_data_model}, has three primary classes of resources. In all cases below, the \texttt{oac} namespace prefix expands to \url{http://www.openannotation.org/ns/}.

\begin{itemize}
	\item The \texttt{oac:Body} of the annotation (node B-1). This resource is the comment, metadata or other information that is created about another resource. The Body can be any Web resource, of any media format, available at any URI. The model allows for exactly one Body per Annotation.
	
	\item The \texttt{oac:Target} of the annotation (node T-1). This is the resource that the Body is about. Like the Body, it can be any URI identified resource. The model allows for one or more Targets per Annotation.
	
	\item The \texttt{oac:Annotation} (node A-1). This resource stands for a document identified by an HTTP URI that describes at least the Body and Target resources involved in the annotation as well as any additional properties and relationships (e.g., dcterms:creator). Dereferencing an annotation's HTTP URI returns a serialization in a permissible RDF format.
	
\end{itemize}

If the Body of an annotation is identified by a dereferencable HTTP URI, as it is the case in Twitter, various blogging platforms, or Google Docs, it can easily be referenced from an annotation. If a client cannot create URIs for an annotation Body, for instance because it is an offline client, they can assign a unique non-resolvable URI (called a URN) as the identifier for the Body node. This approach can still be reconciled with the Linked Data principles as servers that publish such annotations can assign HTTP URIs they control to the Bodies, and express equivalence between the HTTP URI and the URN.



The OAC model also allows to include textual information directly in the annotation document by adding the representation of a resource as plain text to the Body via the \texttt{cnt:chars} property and defining the character encoding using \texttt{cnt:characterEncoding}~\cite{Koch:2009uq}.



\subsection{OAC and Linked Data}
%

Several Linked Data principles have influenced the approach taken by Open Annotation Collaboration. The promotion of a publish/discover approach for handling Annotations as opposed to the protocol-oriented approach taken by Annotea stands out. But also the emphasis on the use (wherever possible) of HTTP URIs for resources involved in Annotations reflects a Linked Data philosophy. Earlier versions of the data model considered an Annotation
to be an OAI-ORE Aggregation \cite{Lagoze:2008fk} and hence a conceptual non-information resource. Community feedback led to a revision of this approach as it was deemed artificial, and the complexities regarding handling the difference between information resource and non-information resource at the protocol level were considered a hindrance to potential adoption.

\subsection{Addressing Media Segments}

Many annotations concern parts, or segments, of resources rather than the entirety of the resource. While simple segments of resources can be identified and referenced directly using the emerging W3C Media Fragment specification~\cite{fragmentsURI:2009fk} or media-type-specific fragment identifiers as defined in RFCs, there are many use cases for segments that currently cannot be identified using this proposal. One example is the annotation of historic maps, where users need to draw polygon shapes around the geographic areas they want to address with their notes. Sanderson et al.~\cite{Sanderson:2011a} describe another use case where annotators express the relationships between images, texts and other resources in medieval manuscripts by means of line segments. For these and other use cases, which require the expression of complex media segment information, the current W3C Media Fragments specification is insufficient.

For this reason, OAC introduces the so-called \texttt{oac:ConstraintTarget} node (CT-1), which constrains the annotation target to a specific segment that is further described by an \texttt{oac:Constraint} (C-1) node. The description of the constraint depends on the application scenario and on the (media) type of the annotated target resource. For example, an SVG path could be used to describe a region within an image.


\subsection{Robust Annotations in Time}

It must be stressed that different agents may create the Annotation, Body and Target at different times. For example, Alice might create an annotation saying that Bob's YouTube video annotates Carol's Flickr photo. Also, being regular Web resources, the Body and Target are likely to have different representations over time. Some annotations may apply irrespective of representation, while others may pertain to specific representations. In order to provide the ability to accurately interpret annotations past their publication, the OAC Data Model introduces three ways to express temporal context. The manner in which these three types of Annotation use the \texttt{oac:when} property, which has a datetime as its value, distinguishes them.

A \emph{Timeless Annotation} applies irrespective of the evolving representations of Body and Target; it can be considered as if the Annotation references the semantics of the resources. For example, an annotation with a Body that says ``This is the front page of CNN'' remains accurate as representations of the Target \url{http://cnn.com/} change over time. Timeless Annotations don’t make use of the \texttt{oac:when} property.

A \emph{Uniform Time Annotation} has a single point in time at which all the resources involved in the Annotation should be considered. This type of Annotation has the \texttt{oac:when} property attached to the Annotation. For example, if Alice recurrently publishes a cartoon at \url{http://example.org/cartoon} that comments on a story on the live CNN home page, an Annotation that has the cartoon as Body and the CNN home page as Target would need to be handled as a Uniform Time Annotation in order to provide the ability to match up correct representations of Body and Target.

A \emph{Varied Time Annotation} has a Body and Target that need to be considered at different moments in time. This type of Annotation uses the \texttt{oac:when} property attached to an \texttt{oac:TimeConstraint} node (a specialization of \texttt{oac:Constraint}) for both Body and Target. If, in the aforementioned cartoon example, Alice would have the habit to publish her cartoon at \url{http://example.org/cartoon} when the mocked article is no longer on the home page, but still use \url{http://cnn.com} as the Target of her Annotation, the Varied Time Annotation approach would have to be used.



This temporal information can be used in the Memento framework to recreate the Annotation as it was intended by reconstructing it with the time-appropriate Body and Target(s). Previous versions of Web resources exist in archives such as the Internet Archive, or within content management systems such as MediaWiki's article history, however they are divorced from their original URI. Memento proposes a simple extension of HTTP in order to connect the original and archived resources. It leverages existing HTTP capabilities in order to support accessing resource versions through the use of the URI of a resource and a datetime as the indicator of the required version. In the framework, a server that host versions of a given resource exposes a TimeGate, which acts as a gateway to the past for a given Web resource. In order to facilitate access to a version of that resource, the TimeGate supports HTTP content negotiation in the datetime dimension. Several mechanisms support discovery of TimeGates, including HTTP Links that point from a resource to its TimeGate(s)~\cite{Sanderson:2010fk}.


\section{Use Cases and Annotation Prototypes}\label{sec:prototypes}

We describe two annotation use cases for which we have implemented early prototypes that publish annotations as Linked Data on the Web following the OAC approach.


\subsection{The OAC/Djatoka Demonstrator}


The OAC/Djatoka Demonstrator implements the current OAC data model for
image resources. It uses the Djatoka image server~\cite{Chute:2008kx}
as its primary platform, which provides panning and zooming
functionality for images using a JPEG 2000 image tiling system.  The
demonstrator enables both creation and viewing of annotations, with
both inline and external resources.

SVG elements can be dynamically created and manipulated using
javascript such that they describe a region of interest.  This
information is then encoded as an \texttt{oac:SvgConstraint} with a UUID URN
identifier, which constrains the full image. The SVG elements may be
resized or repositioned by the user, and scale or translate
respectively with zooming or panning operations.  Instead of part of
the image, the Target may also be one of the other Annotations,
enabling a threaded discussion.

The Body of the annotation is either an external Web resource, 
or a string encoded using
\texttt{cnt:ContentAsText}~\cite{Koch:2009uq}.  The use of external
resources allows the embedded rendering of image, video and audio
within the display, and the re-use of third party hosting services.

The resulting graph is serialized to RDFa and published to online
services such as Blogger. The annotations are then harvested by a
graph database that subscribes to the feeds of known annotators. The
database system~\cite{Sanderson:2006ys} makes the annotations
searchable via the target URI, creation date and content's text.

This prototype affirmed the OAC strategies for strictly adhering to
the Web architecture, allowing any resource to be the Body of an
Annotation rather than just text, and the feasibility of an
environment in which resources are discovered and harvested rather
than transmitted according to a strict protocol, such as in Annotea.

A video of the OAC/Djatoka annotation prototype is available at
\url{http://www.openannotation.org/demos/}.

\subsection{Historic Map Annotation with YUMA}\label{subsec:lemo}


YUMA\footnote{\url{https://github.com/yuma-annotation/}} is an open source annotation framework for the annotation of online multimedia resources. It consists of a server backend and multiple Web clients, each dedicated to a specific media type. At the moment, the YUMA client-suite encompasses clients for the annotation of images, audio, video, and, as a special case of images, maps. Demonstrations of the different YUMA clients are available at: \url{http://dme.ait.ac.at/annotation/}.

\begin{figure}[tbh]
  \centering
  \includegraphics[width=1\columnwidth]{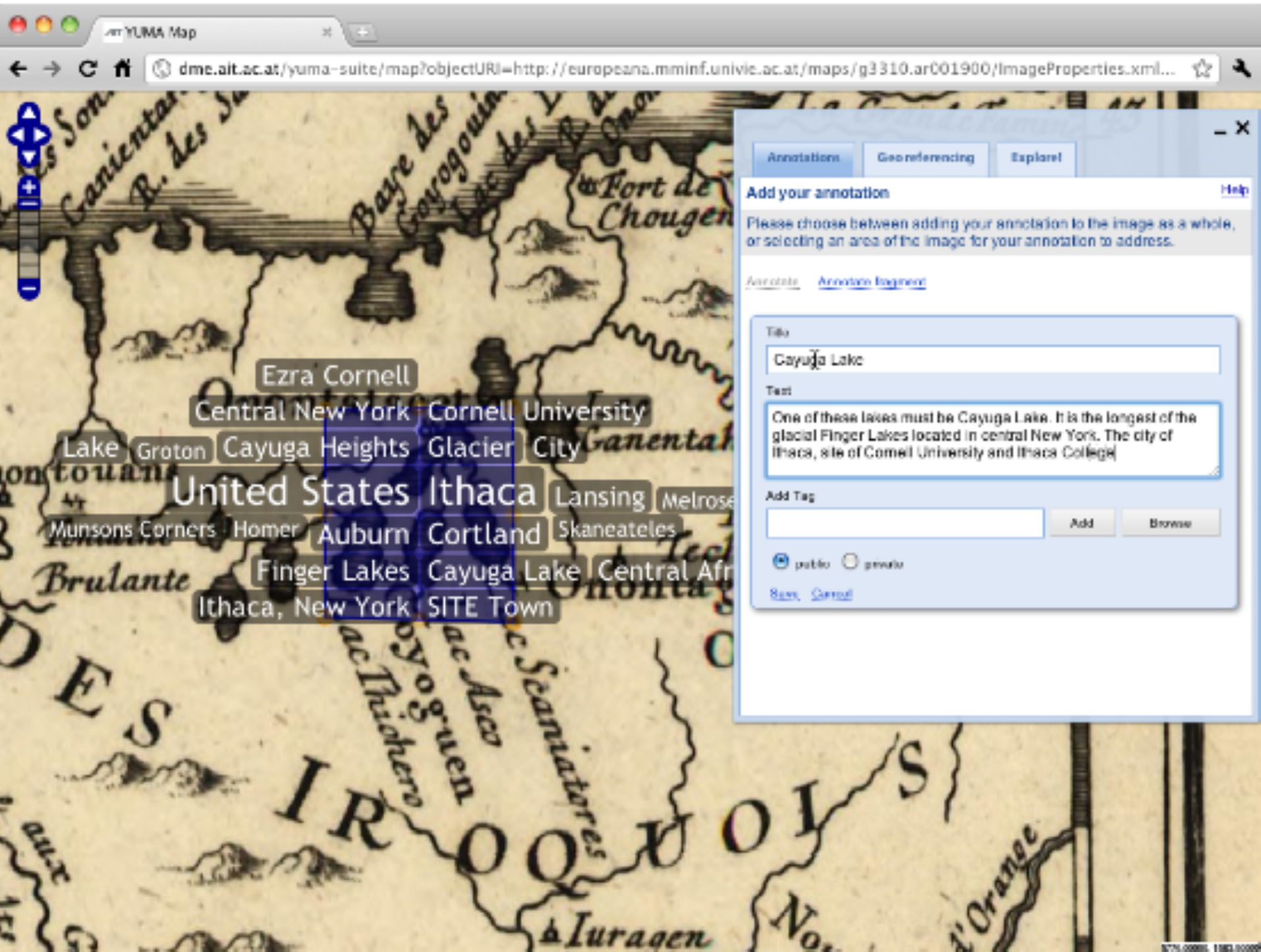}
  \caption{YUMA Map Annotation Tool Screenshot.}
\label{fig:yuma_screenshot}
\end{figure}

The YUMA Map Annotation Tool, which is shown in Figure~\ref{fig:yuma_screenshot}, enables scholars to annotate high-resolution scans of historic maps. It provides similar panning and zooming functionality as the Djatoka annotation service, as well as drawing tools for annotating specific areas on the map. In addition to conventional free-text annotation, YUMA also supports \emph{Semantic Tagging}. When creating or editing annotations, users can make them semantically more expressive by adding references to relevant Linked Data resources on the Web: e.g. links to geographical resources on Geonames or resources from DBpedia. To support users in this task, the tool employs a semi-automatic approach. Based on the annotated geographical area, as well as on an analysis of the annotation text, potentially related Linked Data resources are proposed automatically in the form of a tag cloud. The user is prompted to verify the proposed resources, or can simply ignore them. For those resources that have been user-verified, the system adds the link to the annotation. Additionally, the system dereferences the resource and stores relevant properties as part of the annotation metadata: e.g. alternative language labels, spelling variants, or geographical coordinates. This information can later be exploited to facilitate advanced search functionalities such as multilingual, synonym, or geographical search~\cite{Simon:2011fk}.

\begin{figure}[tbh]
  \centering
  \includegraphics[width=1\columnwidth]{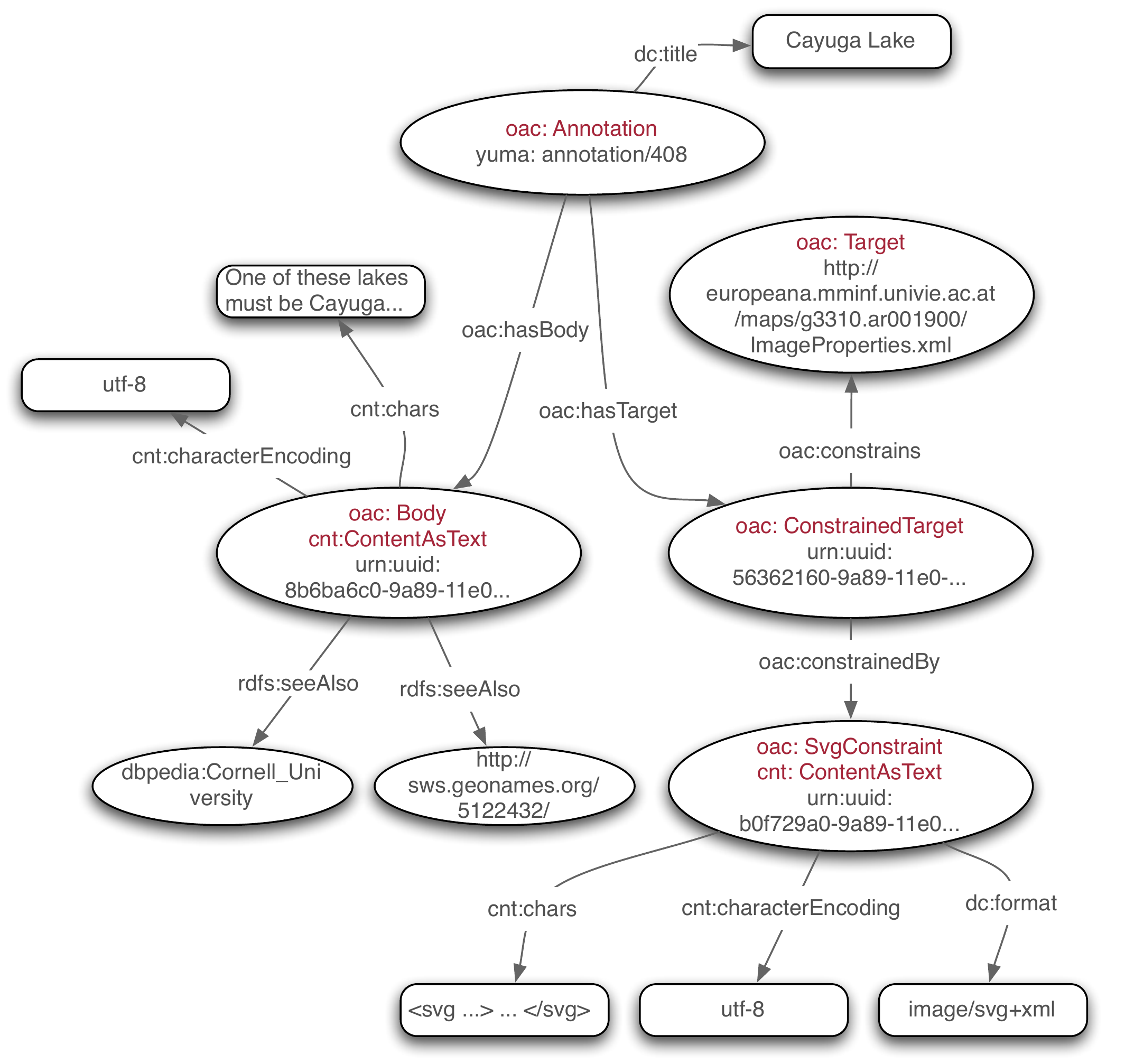}
  \caption{Sample YUMA OAC Annotation.}
\label{fig:yuma_oac}
\end{figure}

The YUMA server backend exposes annotations as Linked Data on the Web following the OAC model. As illustrated with an example annotation in Figure~\ref{fig:yuma_oac}, each Annotation resource has its own dereferencable HTTP URI. The textual annotation Body is embedded directly into the Annotation document and has a unique non-resolvable URI (URN) as identifier. The annotation Body is about a Web resource - in this case a high-resolution zoomable image, published as a Zoomify\footnote{\url{http://zoomify.com/}} tileset. The tileset is identified by its XML metadata descriptor file, which acts as the annotation Target. The annotated region within the zoomable image is defined by means of an \texttt{oac:SvgConstraint} resource, which allows us to add an SVG snippet expressing the boundaries of that region, to that resource.

Because of the limited expressiveness of \texttt{oac:Body} in the Alpha3 OAC specification, user-contributed links are currently loosely attached to the Body. This results in a loss of semantics because some of the proposed links annotate certain text segments or named entities in the annotation Body. We plan to change this representation as soon as OAC provides means for representing structured information in annotation Bodies.



%
%
%

\section{Related Work}\label{sec:related_work}


Annotations have a long research history, and unsurprisingly the research perspectives and interpretations of what an \emph{annotation} is supposed to be vary widely. Agosti et al.~\cite{Agosti:2007uq} provide a comprehensive study on the contours and complexity of annotations. A representative discussion on how annotations can be used in various scholarly disciplines in given by Bradley~\cite{Bradley:2008kx}. He describes how annotations can support interpretation development by collecting notes, classifying resources, and identifying novel relationships between resources.

Besides Annotea other annotation models have been proposed: \cite{Arndt:2007uq} built the MPEG-7 compliant COMM Ontology. OAC, in contrast, is a resource-centric annotation model, which is more light-weight because it doesn't have a background in the automated feature extraction and representation. The M3O Ontology~\cite{Saathoff:2010vn} allows the integration of annotations with SMIL and SVG documents. OAC, in contrast, treats annotations as first-class resources on the Web, which would not be part but \emph{about} a presentation.

Early related work on the issue of describing segments in multimedia resources can be traced back to research on linking in hypermedia documents~\cite{Hardman:1994zr}. For describing segments using a non-URI based mechanism one can use MPEG-7 Shape Descriptors (cf.~\cite{Nack:1999ly}) or terms defined in a dedicated multimedia ontology. SVG~\cite{svg:2003bh} and MPEG-21~\cite{ISO/IEC:2006qf} introduced XPointer-based URI fragment definitions for linking to segments in multimedia resources.

\section{Conclusions and Future Work}\label{sec:summary}

We apply a generic and Web-centric conception to the various facets annotations can have and regard an annotation as association created between one \emph{body} resource and other \emph{target} resources, where the body must be somehow \emph{about} the target. This conception lead to the specification of the OAC model, which originates from activities in the Open Annotation Collaboration and aims at building an interoperable environment for publishing annotations on the Web. We also presented and provided pointers to two prototypes that currently implement the OAC specification.

The OAC specification is currently in Alpha3 stage and our future work will focus on the following issues: support for structured bodies that go beyond resource referencing and string-literal representation, extension mechanisms for addressing complex media segments in various media types, and processing of constraints.

We will further pursue the integration of the OAC segment identification approach with the W3C Media Fragment Identification mechanism. Since it is hardly possible to address all possible segment shapes in a fragment identification specification, we propose an additional fragment key/value pair for the spatial dimension, which enables fragment identification by reference in W3C Media Fragment URIs. The key could be \emph{ptr}, \emph{ref} or something similar and the value a URI. The URI points to a resource, which provides further information about the properties of the spatial region/segment. We suggested this to the Media Fragment Working Group and hope that this issue will be addressed in future W3C Media Fragments URI recommendations.

As a final result, we expect a light-weight annotation model that is straightforward to use in basic annotation use cases but provides extension points for more complex annotation scenarios.

\section*{Acknowledgment}

The work has partly been supported by the European Commission as part of the eContentplus program (EuropeanaConnect) and by a Marie Curie International Outgoing Fellowship within the 7th Europeana Community Framework Programme". The development of OAC is funded by the Andrew W. Mellon foundation.



\bibliographystyle{IEEEtran}
\bibliography{IEEEabrv,references}
%
%
%

\end{document}